\documentclass[journal=nalefd,manuscript=letter]{achemso}

\usepackage{graphicx}
\usepackage{amssymb}
\usepackage{amsmath}
\usepackage{bm}
\usepackage[breaklinks]{hyperref}
\usepackage[all]{hypcap}
\usepackage{siunitx}
\hypersetup{
    plainpages=false,
    bookmarks=false,         
    unicode=false,          
    pdfborder={0 0 0}
    pdftoolbar=true,        
    pdfmenubar=true,        
    pdffitwindow=false,     
    pdfstartview={FitH},    
    pdftitle={},    
    pdfauthor={},     
    pdfproducer={LNCMI-CNRS-UGA-UPS-INSA}, 
    pdfnewwindow=true,      
    linktoc=section,
    colorlinks=true,       
    linkcolor=blue,          
    citecolor=red,        
    filecolor=magenta,      
    urlcolor=blue           
}

\author{Alessandro Surrente}
\affiliation{Laboratoire National des Champs Magn\'etiques Intenses,
UPR 3228, CNRS-UGA-UPS-INSA, Grenoble and Toulouse, France}

\author{Dumitru Dumcenco}
\affiliation{Electrical Engineering Institute and Institute of
Materials Science and Engineering, \'{E}cole Polytechnique
F\'{e}d\'{e}rale de Lausanne, CH-1015 Lausanne, Switzerland}

\author{Zhuo Yang}
\affiliation{Laboratoire National des Champs Magn\'etiques Intenses,
UPR 3228, CNRS-UGA-UPS-INSA, Grenoble and Toulouse, France}

\author{Agnieszka Kuc}
\affiliation{University of Leipzig, Wilhelm Ostwald Institute of
Physical and Theoretical Chemistry Leipzig, Saxony, Germany}
\affiliation{School of Engineering and Science, Jacobs University
Bremen, Campus Ring 1, 28759 Bremen, Germany}

\author{Yu Jing}
\affiliation{University of Leipzig, Wilhelm Ostwald Institute of
Physical and Theoretical Chemistry Leipzig, Saxony, Germany}
\affiliation{School of Engineering and Science, Jacobs University
Bremen, Campus Ring 1, 28759 Bremen, Germany}

\author{Thomas Heine}
\affiliation{University of Leipzig, Wilhelm Ostwald Institute of
Physical and Theoretical Chemistry Leipzig, Saxony, Germany}
\affiliation{School of Engineering and Science, Jacobs University
Bremen, Campus Ring 1, 28759 Bremen, Germany}

\author{Yen-Cheng Kung}
\affiliation{Electrical Engineering Institute and Institute of
Materials Science and Engineering, \'{E}cole Polytechnique
F\'{e}d\'{e}rale de Lausanne, CH-1015 Lausanne, Switzerland}

\author{Duncan K.\ Maude}
\affiliation{Laboratoire National des Champs Magn\'etiques Intenses,
UPR 3228, CNRS-UGA-UPS-INSA, Grenoble and Toulouse, France}

\author{Andras Kis}
\affiliation{Electrical Engineering Institute and Institute of
Materials Science and Engineering, \'{E}cole Polytechnique
F\'{e}d\'{e}rale de Lausanne, CH-1015 Lausanne, Switzerland}

\author{Paulina Plochocka}
 \email{paulina.plochocka@lncmi.cnrs.fr}
\affiliation{Laboratoire National des Champs Magn\'etiques Intenses,
UPR 3228, CNRS-UGA-UPS-INSA, Grenoble and Toulouse, France}

\date{\today}

\title[]
{Defect healing and charge transfer mediated valley polarization in
MoS$_2$/MoSe$_2$/MoS$_2$ trilayer van der Waals heterostructures}


\begin{document}
\begin{tocentry}

    \includegraphics[width=1.0\linewidth]{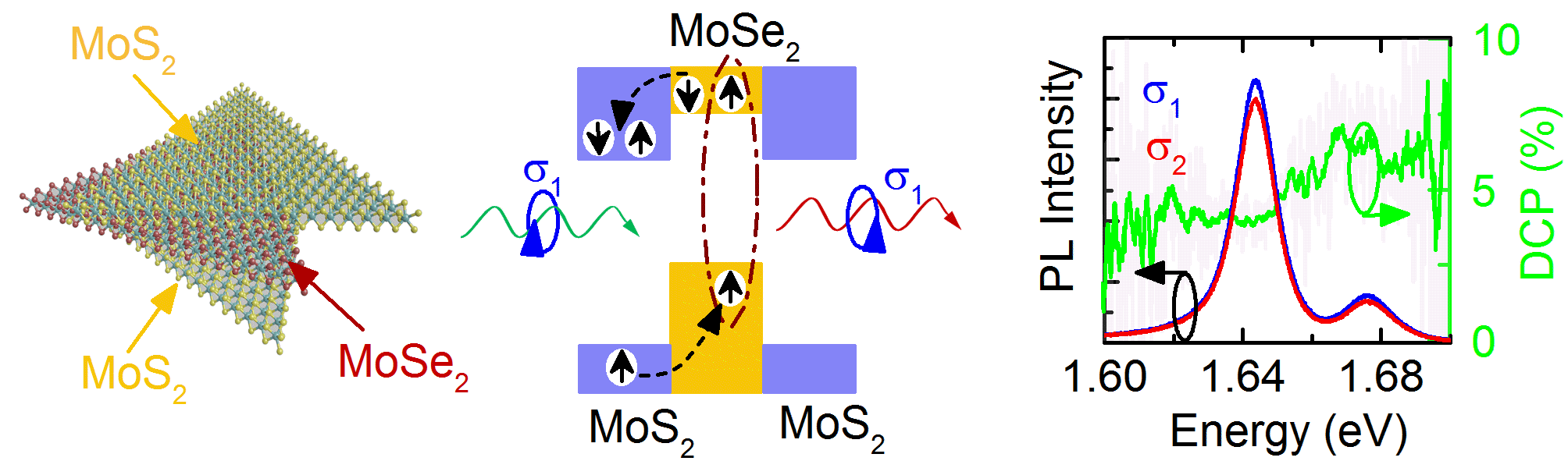}

\end{tocentry}

\begin{abstract}
Monolayer transition metal dichalcogenides (TMDC) grown by chemical
vapor deposition (CVD) are plagued by a significantly lower optical
quality compared to exfoliated TMDC. In this work we show that the
optical quality of CVD-grown MoSe$_2$ is completely recovered if the
material is sandwiched in MoS$_2$/MoSe$_2$/MoS$_2$ trilayer van der
Waals heterostructures. We show by means of density-functional
theory that this remarkable and unexpected result is due to defect
healing: S atoms of the more reactive MoS$_2$ layers are donated to
heal Se vacancy defects in the middle MoSe$_2$ layer. In addition,
the trilayer structure exhibits a considerable charge-transfer
mediated valley polarization of MoSe$_2$ without the need for
resonant excitation. Our fabrication approach, relying solely on
simple flake transfer technique, paves the way for the scalable
production of large-area TMDC materials with excellent optical
quality.

\end{abstract}



\vspace{3cm}
Monolayer transition metal dichalcogenides (TMDC) have a direct
bandgap situated in the visible range, which makes them ideal
building blocks for novel electronic and optoelectronic devices
\cite{Splendiani10,mak2010atomically,radisavljevic2011single,kuc2011influence,lopez2013ultrasensitive,baugher2014optoelectronic,Ross2014,Pospischil2014,Koppens2014,mak2016photonics}.
The bandgap of monolayer TMDCs occurs at the inequivalent (but
degenerate) K and K' points of the hexagonal Brillouin zone. The
broken inversion symmetry of a TMDC monolayer combined with the time
reversal symmetry imposes opposite magnetic moments at the K and K'
valleys. This in turn determines the characteristic circular
dichroism exhibited by these materials, wherein each valley can be
addressed separately with circularly polarized light of a given
helicity \cite{cao2012valley,zeng2012valley,mak2012control}.
Additionally, optical spectra are influenced by the strong
spin-orbit coupling, which lifts the degeneracy of band states at
the valence band edges, resulting in well-resolved A and B
resonances, as observed in reflectivity or absorption spectra
\cite{mak2010atomically,li2014measurement,he2014tightly,mak2013tightly}.
The interplay of spin-orbit coupling with broken inversion symmetry
and time reversal symmetry locks the valley and spin degrees of
freedom, making TMDC attractive candidates for valleytronics
\cite{xiao2012coupled}. The spin-valley index locking along with the
large distance in the momentum space between K and K' valleys
preserves the valley polarization observed in the degree of circular
polarization (DCP) in helicity resolved photoluminescence
emission\cite{jones2013optical,plechinger2015identification,wang2015giant,srivastava2015valley,Zhu2011}.

Applications require a scalable fabrication platform providing high-quality large-area monolayer TMDC. Unfortunately, the most
promising approach today, namely chemical vapor deposition (CVD)
growth\cite{zhan2012large,dumcenco2015large,wang2014chemical,zhang2013controlled,huang2013large} struggles to compete with
exfoliated TMDC in terms of sample quality. Low temperature PL spectroscopy of CVD-grown MoS$_2$ and MoSe$_2$ reveals broad
emission from defect bound excitons, which is significantly more intense than the free exciton
peak \cite{chang2014monolayer,han2016photoluminescence,li2016suppression} and is related to chalcogen vacancies induced during
the CVD growth \cite{han2016photoluminescence,li2016suppression}.

Here, we demonstrate a novel approach to neutralize the intrinsic
defects of CVD-grown TMDCs, using flake transfer tools routinely
employed in the fabrication of van der Waals heterostructures
\cite{chiu2014spectroscopic,wang2016interlayer,tongay2014tuning}. We
investigate the optical properties of trilayer stacks composed of
external CVD-grown MoS$_2$ flakes \cite{dumcenco2015large} as
capping layers and an internal CVD-grown MoSe$_2$ flake which has a
smaller bandgap \cite{mitioglu2016magnetoexcitons,kang2013band}.
Remarkably, this fabrication approach strongly suppresses the
localized exciton emission in MoSe$_2$, yielding a low temperature
PL comparable to that observed in mechanically exfoliated samples.
This striking result can be understood from density functional
theory (DFT), which suggests that the more reactive MoS$_2$ donates
chalcogen atoms to heal vacancy defects in MoSe$_2$. Incorporating
MoS$_2$ into the trilayer heterostructure furthermore allows us to
demonstrate a new way to introduce valley polarization in MoSe$_2$.
Due to the type II band alignment in TMDC heterojunctions
\cite{kang2013band}, a significant charge transfer is observed in
these systems
\cite{hong2014ultrafast,ceballos2014ultrafast,wang2016interlayer}.
Our results show that spin of the hole is conserved upon charge
transfer from MoS$_2$ to MoSe$_2$ after excitation in resonance with
MoS$_2$ A exciton. This leads to non-zero steady state valley
polarization in MoSe$_2$, which has never been observed before under
non-resonant excitation \cite{wang2015polarization,kioseoglou2016optical,baranowski2017dark}.

\section{Defect healing}
\begin{figure}[h]
\includegraphics[width=1.0\linewidth]{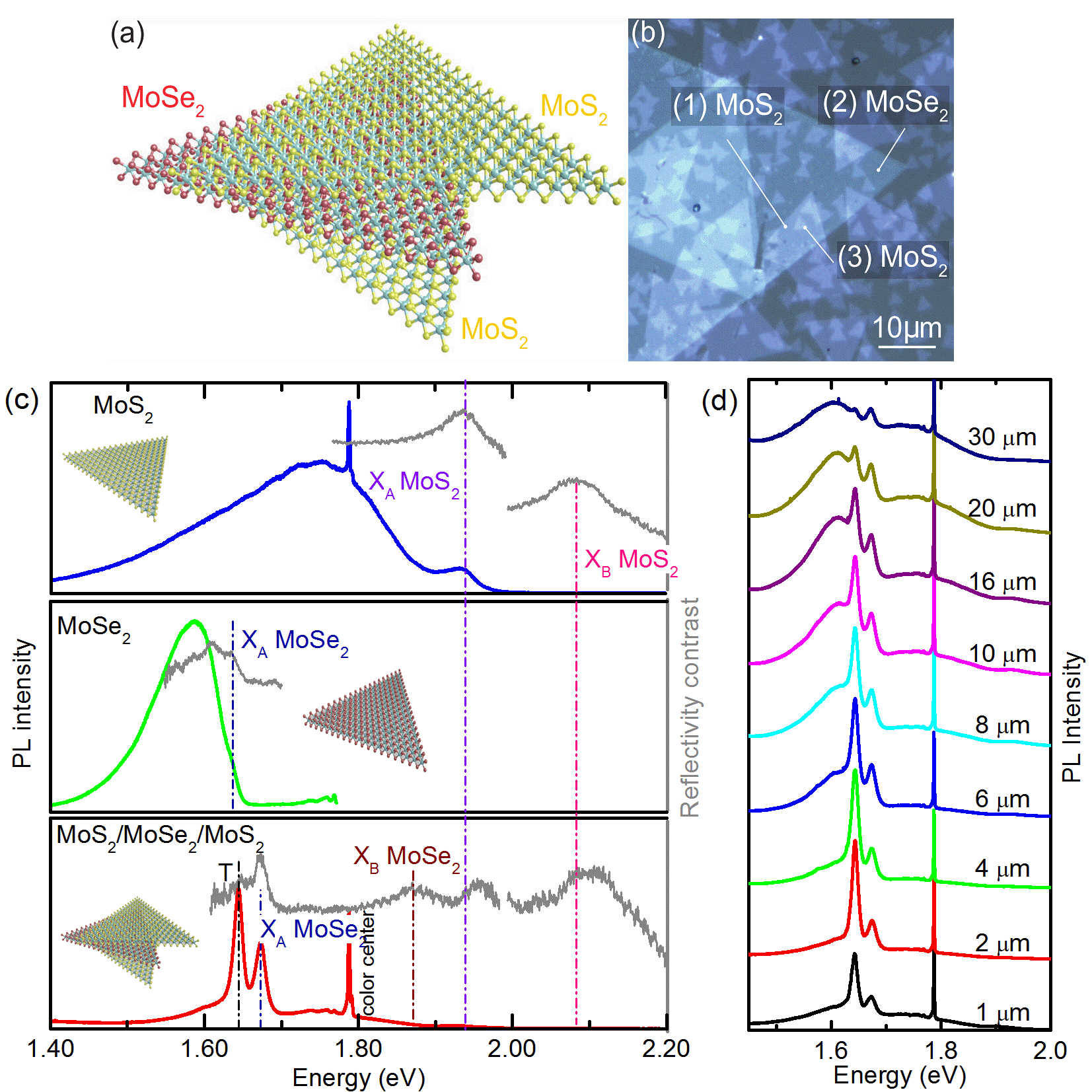}
\caption{(a) Ball and stick model of MoSe$_2$ layer sandwiched
between two MoS$_2$ flakes. (b) Optical micrograph of a
representative transfer area. (c) Low temperature \si{\micro}PL spectra (colored curves) and reflectivity contrast (grey curves) of
as-grown MoS$_2$, MoSe$_2$ monolayers and MoS$_2$/MoSe$_2$/MoS$_2$
trilayer stack. (d) PL spectra of trilayer stack measured using
different spot sizes. The spectra are vertically offset for clarity.
X$_{\text{A}}$ indicates A exciton and T denotes the trion (charged
A exciton).} \label{fig:TrilayerPL}
\end{figure}

The sample with MoS$_2$/MoSe$_2$/MoS$_2$ trilayer stacks and micrograph of a representative transfer area are schematically shown
in Fig.\,\ref{fig:TrilayerPL}(a) and (b). Low temperature microPL (\si{\micro}PL) spectroscopy has been used to characterize
\emph{as-grown} CVD samples on the sapphire substrate (prior to any transfer process) and the trilayer MoS$_2$/MoSe$_2$/MoS$_2$
stack. All spectra shown in Fig.\,\ref{fig:TrilayerPL}(c) and (d) were measured under nominally identical conditions with an
excitation power of \SI{100}{\micro\W} (see Fig.\ S1 in Supporting Information (SI) for power dependent
\si{\micro}PL spectra). The \emph{as-grown} MoS$_2$ and MoSe$_2$ monolayers both show a broad PL feature (full width at half
maximum, FWHM, of \SI{258}{\milli\eV} and \SI{106}{\milli\eV}, respectively) related to emission from excitons bound to defect or
charge impurity states \cite{chang2014monolayer,tongay2013defects,han2016photoluminescence,li2016suppression}. A-exciton emission
(labeled X$_\text{A}$ in Fig.\ \,\ref{fig:TrilayerPL}(c)) is seen only as a weak peak or shoulder at higher energies.
These assignments are confirmed by reflectivity contrast measured on the same spot. Reflectivity contrast is
defined as $\Delta R/R_{\text{s}}=(R-R_{\text{s}})/R_{\text{s}}$, where $R$ is the reflectivity spectrum measured on the sample
and $R_{\text{s}}$ denotes the reflectivity spectrum measured on the substrate. In the case of transparent substrates such as
sapphire, reflectivity contrast is proportional to the absorption of the sample \cite{rigosi2015probing}. The reflectivity
contrast spectrum of as-grown MoS$_2$ (see upper panel of Fig.\ \ref{fig:TrilayerPL}(c)) consists of a main peak at
\SI{1.937}{\eV} and a higher energy, weaker feature at \SI{2.085}{\eV}, related to the spin-orbit split B exciton. The energy
difference of \SI{148}{\milli\eV} corresponds very well to theoretically predicted spin-orbit splitting of the valence band
\cite{ramasubramaniam2012large} and is very similar to the splitting determined with transmission measurements on similar samples
\cite{mitioglu2016magnetoexcitons}. The reflectivity contrast spectrum of as grown MoSe$_2$, shown in the central panel of Fig.\
\ref{fig:TrilayerPL}(c) has a peak at \SI{1.637}{\eV}, which corresponds well to the high energy shoulder of the \si{\micro}PL
spectrum and hence is assigned to A exciton. As in our previous study \cite{mitioglu2016magnetoexcitons}, we are unable to
resolve the B exciton in reflectivity contrast measurements on as grown MoSe$_2$.

The optical properties of the trilayer stack are dramatically improved (bottommost panel in Fig.\,\ref{fig:TrilayerPL}(c)). The
most striking difference is the nearly total suppression of (i) emission from defect bound excitons, and (ii) significant
quenching of both MoS$_2$ PL, barely seen as a weak peak at \SI{1.929}{\eV}, and of MoSe$_2$ PL
(overall integrated intensity decrease by two and one orders of magnitude, respectively, see Fig.\ S2 in SI).
The measured PL spectrum is dominated by narrow free neutral (X$_\text{A}$) and charged exciton (T) emission in MoSe$_2$
(low power FWHM of \SI{11}{\milli\eV} and \SI{10}{\milli\eV}, respectively, see Fig.\ S1 in SI), approaching the
quality of exfoliated WSe$_2$ embedded in boron nitride (FWHM $\sim \SI{10}{\milli\eV}$) \cite{withers_wse2_2015}. The free
exciton emission overlaps with a weak broad background emission from the sapphire substrate (the narrow peak just below
\SI{1.8}{\eV} corresponds to the emission from a color center in sapphire). The highest energy peak in the
reflectivity contrast spectrum of the trilayer (see Fig.\ \ref{fig:TrilayerPL}(c), bottom panel), blue shifted with respect to
the A exciton peak of MoS$_2$ by \SI{145}{\milli\eV} is attributed to the B exciton of MoS$_2$. The peak related to MoS$_2$ A
exciton in the reflectivity contrast spectrum is blue shifted by \SI{27}{\milli\eV} with respect to the corresponding PL peak.
This Stokes shift has been attributed to the presence of a high doping level in MoS$_2$ \cite{mak2013tightly}, also present in our layers \cite{dumcenco2016high}. On the low energy side of the
\si{\micro}PL spectrum, the two distinct peaks at \SI{1.674}{\eV} and \SI{1.645}{\eV} are assigned to the A exciton and to the
trion of MoSe$_2$, respectively. The trion binding energy of \SI{29}{\milli\eV} is very similar to that reported in other studies
on MoSe$_2$ \cite{han2016photoluminescence,ross2013electrical}.

The high optical quality of the MoSe$_2$ embedded in the heterostructure enabled us to resolve an additional peak
at \SI{1.87}{\eV} in the reflectivity contrast spectrum, assigned to the B exciton of MoSe$_2$
\cite{ramasubramaniam2012large,arora2015exciton}. The vastly improved optical properties suggest a defect healing process, in
which the contact with MoS$_2$ is enough to drastically reduce the number of defects in MoSe$_2$. The quenching of the
intralayer emission \cite{wang2016interlayer,ceballos2014ultrafast} in the trilayer is manifestation of a fast
charge transfer mechanism \cite{hong2014ultrafast,yu2014equally} related to the type II band alignment in MoS$_2$/MoSe$_2$
heterostructures \cite{kang2013band}. The weak luminescence of MoS$_2$ is consistent with a background n doping
\cite{dumcenco2016high} of the as grown layers and with an additional charge transfer after the formation of the
heterostructure. We assign the brighter emission from MoSe$_2$ (in the trilayer) to the hole transfer to an intrinsically n doped
material and to the defect healing effect, combined with luminescence resulting from higher energy states, similar to hot
luminescence of direct exciton in multilayer MoSe$_2$ \cite{tongay2012thermally,tonndorf2013photoluminescence}. The long range
optical uniformity of the trilayer stacks has been monitored by acquiring PL with a gradually defocused excitation beam. The
acquired PL spectra are displayed in Fig.\,\ref{fig:TrilayerPL}(d). For an excitation spot size of
\SI{10}{\micro\m}, the defect emission remains strongly suppressed. For larger spot sizes, a broad low-energy
peak starts to emerge, probably due to defect related emission in MoSe$_2$ in areas which are not fully capped. These
measurements are a proof of concept, demonstrating that this approach, when optimized, should enable the fabrication of large
area CVD-grown heterostructures with excellent optical quality.

Previous attempts at improving the optical properties of CVD-grown MoSe$_2$ using HBr treatment \cite{han2016photoluminescence}
or the isoelectronic impurity substitution \cite{li2016suppression} have met with only partial success: Impurity-bound excitons
still remained the most prominent component of the emission spectrum. Low temperature PL spectra consisted in broad features,
wherein free exciton emission could be identified only after fitting. The optical properties of exfoliated
MoS$_2$ have been improved by superacid treatment \cite{cadiz2016well}. In our case, the simple act of bringing MoSe$_2$ in
intimate physical contact with MoS$_2$, a procedure that can be performed after growth and does not require any chemical
functionalisation, results in a virtually complete suppression of emission from the impurity-bound states and a spectrum in which
the trion and exciton resonances can be clearly resolved.


Our defect healing hypothesis is further supported by the results obtained from DFT simulations. We have calculated the gain of
energy in a heterobilayer MoSe$_2$/MoS$_2$ using two models: A single Se vacancy in the MoSe$_2$ layer (Model 1) and a single S
vacancy in the MoS$_2$ layer together with a single S substitution in the MoSe$_2$ layer (Model 2). This corresponds to the
transfer energy of a S atom from pristine MoS$_2$ to heal a defect in MoSe$_2$. We observe a significant energy gain of
18\,kJ\,mol$^{-1}$ (180\,meV) per S transfer from MoS$_2$ to the MoSe$_2$ defect, which shows that defect healing in MoSe$_2$ by
MoS$_2$ is thermodynamically favored. The formation energy of a S vacancy in a MoS$_2$ monolayer has been
theoretically estimated in the \SI{1.3}{\eV}--\SI{1.5}{\eV} range \cite{noh2014stability,komsa2015native}. We consider these as
upper bounds for the energy barrier of the transfer of a S atom to fill a Se vacancy in MoSe$_2$, because this is not a static
process but a transfer between two neighbouring layers.

\begin{figure}[h]
\includegraphics[width=12 cm]{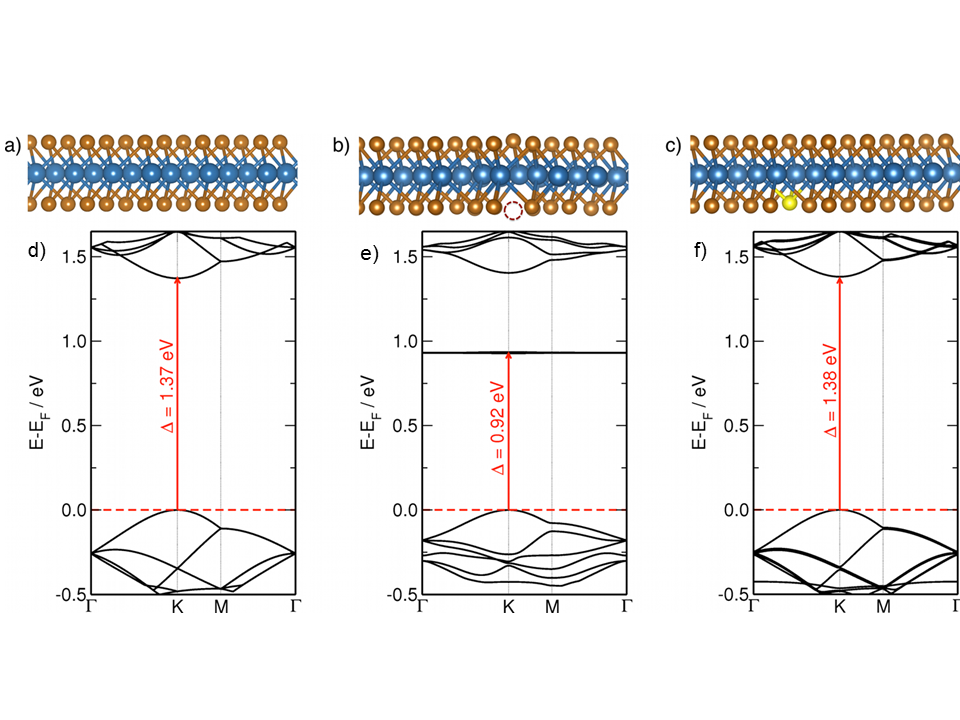}
\caption{(a)--(c) Atomic representation of a perfect MoSe$_2$ monolayer, the MoSe$_2$ monolayer with 1 Se vacancy, and the
MoSe$_2$ monolayer with 1 S substitution, respectively. (d)--(f) Calculated corresponding band structures. Fermi level
(horizontal dashed lines) is shifted to the top of the valence band. Fundamental bandgaps are indicated.}
\label{fig:DFTbandstructure}
\end{figure}

We have also calculated the band structures of a perfect MoSe$_2$ monolayer, MoSe$_2$ monolayer with one Se vacancy, and MoSe$_2$
monolayer with one Se$\longrightarrow$S substitution. We observe that Se vacancies introduce strongly localized states in the
bandgap of MoSe$_2$, 0.92\,eV above the top of the valence band (see Fig.\ \ref{fig:DFTbandstructure}). These are dispersionless
and act as trap centers. Healing the Se vacancy with S substitution restores the band structure of a nearly perfect MoSe$_2$
monolayer. In the studied $5 \times 5$ supercell model (2$\%$ Se$\longrightarrow$S substitutions), the bandgap increases by only
1 meV.

\section{Charge transfer mediated valley polarization}

\begin{figure}
\centering
\includegraphics[width=0.75\linewidth]{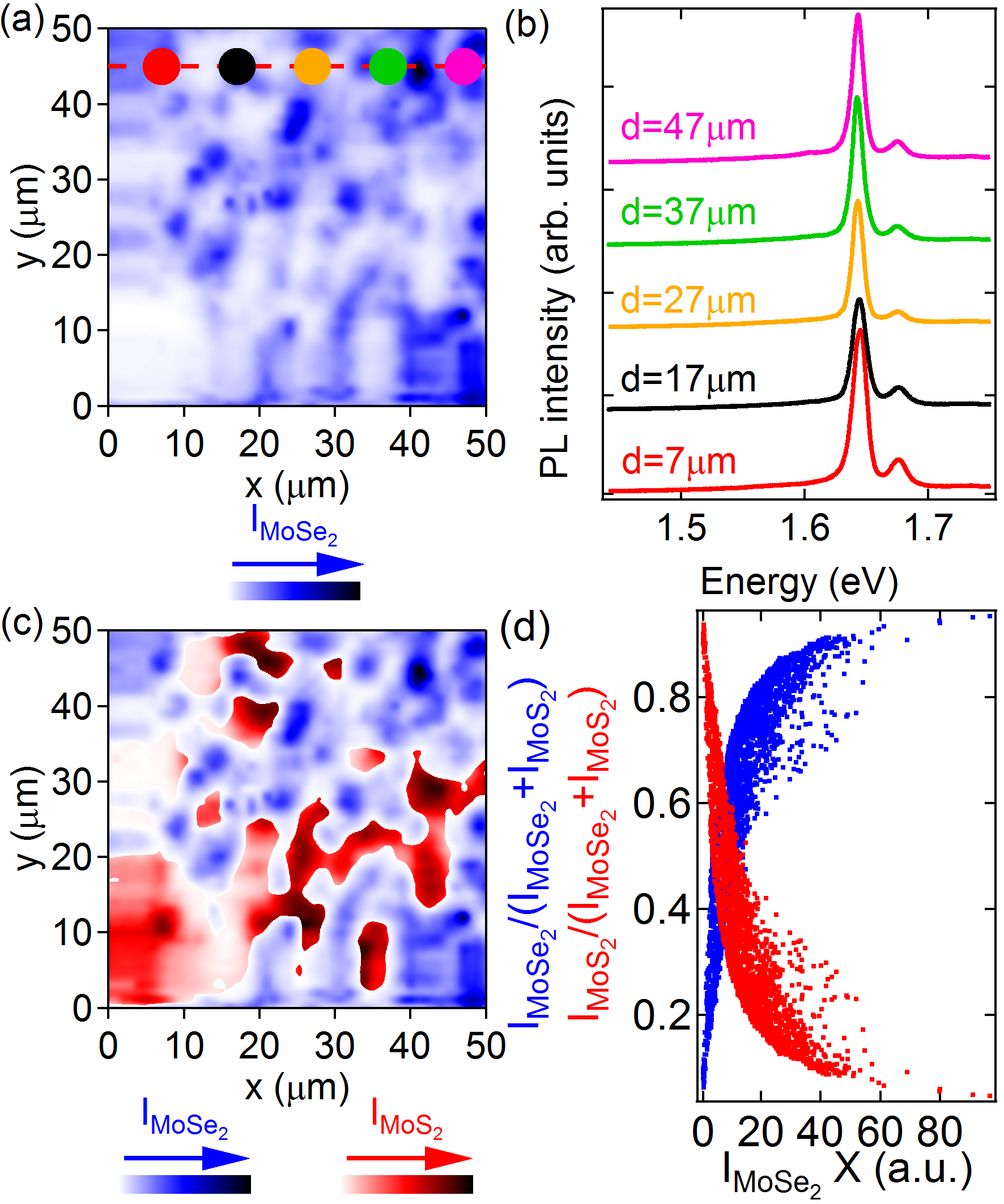}
\caption{(a) Spatial map of the integrated intensity of A exciton of encapsulated MoSe$_2$. The dashed line schematically
illustrates the direction along which the spectra of panel (b) have been extracted. The colored circles mark the position within
the map of the spectra having the corresponding color in panel (b). (b) \si{\micro}PL spectra extracted at positions marked by
circles of the corresponding color in panel (a). The spectra are vertically offset for clarity. (c) Mapping of the integrated
intensity of MoS$_2$ X$_{\text{A}}$ overlaid on the same spatial map of MoSe$_2$ X$_{\text{A}}$. (d) Integrated intensity of
MoSe$_2$ X$_{\text{A}}$ and MoS$_2$ X$_{\text{A}}$ as a function of the intensity of MoSe$_2$ X$_{\text{A}}$.}
\label{fig:MappingTimeResolved}
\end{figure}

First signatures of charge transfer between the layers are seen in the comparison between PL intensities of as
grown layers and trilayers, shown in Fig.\ S2 of SI. Additional insights is provided by the spatial correlation of the PL
intensity of the MoS$_2$ and MoSe$_2$. The integrated spatial map of the MoSe$_2$ A exciton is shown in
Fig.\,\ref{fig:MappingTimeResolved}(a). The signal is particularly intense at positions where the overlap between the three
layers is good and the material does not have a large number of defects. This implies that the bright spots do not necessarily
have a triangular shape. This map provides an additional opportunity to demonstrate the high degree of uniformity
of the emission of MoSe$_2$ incorporated in a heterostructure, by extract \si{\micro}PL spectra. We show in Fig.\
\ref{fig:MappingTimeResolved}(b) five \si{\micro}PL spectra measured at \SI{10}{\micro\m} distance from one another. The spectra
have been normalized by the integration time. We note that emission from defect states is consistently absent in the five spectra
and the similar line shape points to a good uniformity of the emission over the full mapped area. In
Fig.\,\ref{fig:MappingTimeResolved}(c), we overlay the intensity map of MoS$_2$ with that of MoSe$_2$, forcing the areas having
the lowest signal from the latter to be transparent. These areas correspond to zones where the signal from MoS$_2$ is highest. We
quantify the observed intensity (anti)correlation by plotting in Fig.\,\ref{fig:MappingTimeResolved}(d) the ratio between the
intensity of A exciton in one material normalized by the total emission of both materials
[$I_{\text{MoX}_2}/(I_{\text{MoS}_2}+I_{\text{MoSe}_2}$), where $X$ = S or Se]. It can be noted that when the emission of
MoSe$_2$ becomes more pronounced, the emission of MoS$_2$ decreases correspondingly. This is fully consistent
with charge transfer. In positions where the three layers overlap efficiently, charge transfer induces a
quenching of the MoS$_2$ PL. At the same time, the PL from MoSe$_2$ is particularly intense at these positions owing to efficient
defect healing, but still weaker than in as grown MoSe$_2$ samples (see Fig.\ S2 in SI).

\begin{figure}[h]
\includegraphics[width=14 cm]{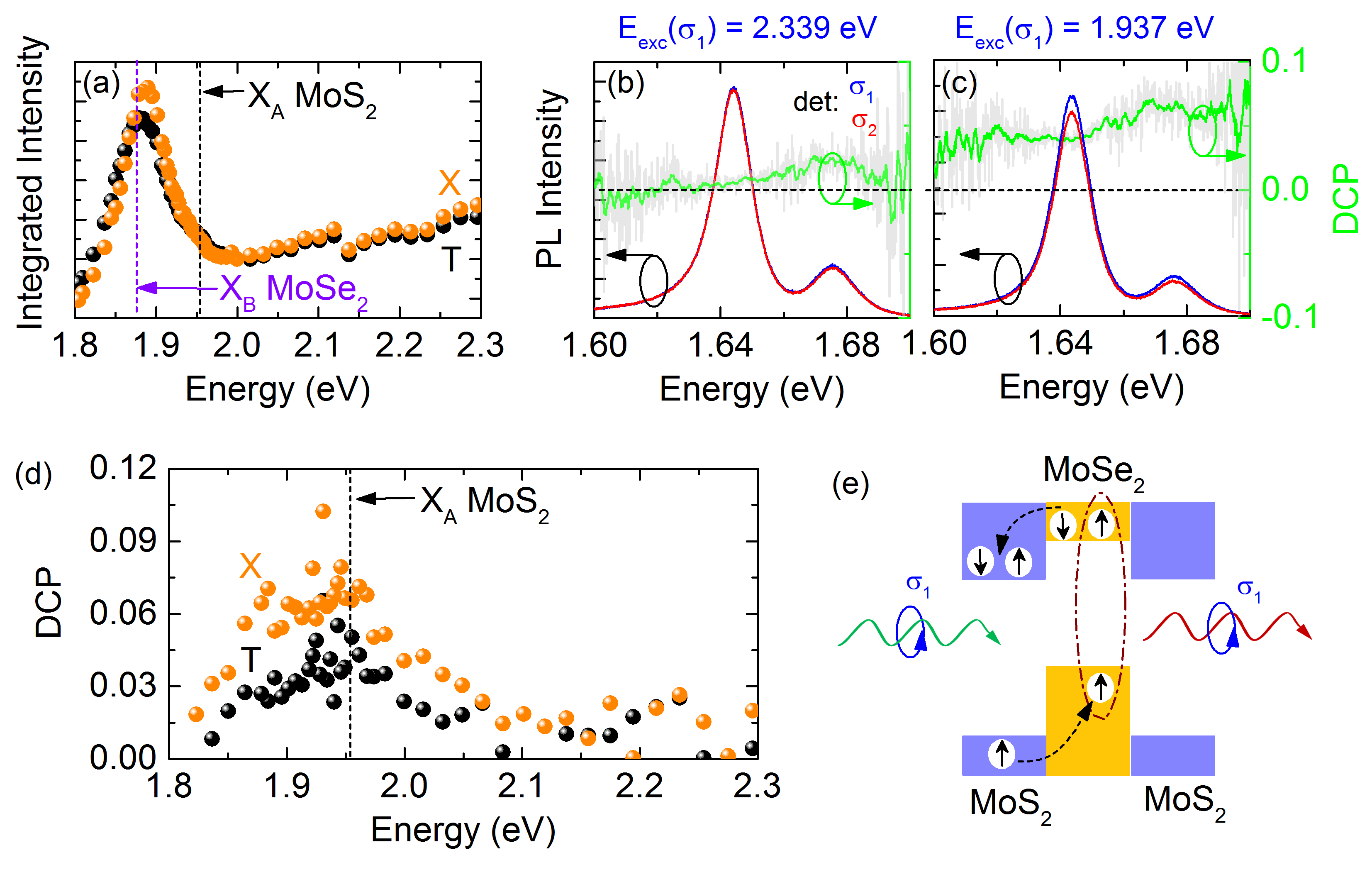}
\caption{(a) Normalized integrated intensity of MoSe$_2$
X$_{\text{A}}$ as a function of the excitation energy. Polarization
resolved \si{\micro}PL spectra and DCP of MoSe$_2$ excited (b) far off
MoS$_2$ X$_{\text{A}}$ resonance and (c) on resonance with
MoS$_2$ X$_{\text{A}}$. (d) Integrated DCP as a function of the
excitation energy. (e) Schematic illustration of spin transfer under
excitation resonant with MoS$_2$ X$_\text{A}$.}
\label{fig:DOP_PLE}
\end{figure}

Photoluminescence excitation (PLE) spectroscopy is a useful tool to investigate more thoroughly the charge transfer processes. We
focus our measurements on MoSe$_2$ incorporated in a trilayer stack. In Fig.\,\ref{fig:DOP_PLE}(a), we show the PLE measured
without making use of polarization optics. Both PLEs of the A exciton and trion have been normalized by the weakest intensity
measured at an excitation energy of 1.999 eV in order to be able to compare the enhancement effects for the exciton and trion.
The integrated intensities of both peaks show a pronounced maximum when the excitation energy is close to the resonance with the
B exciton of MoSe$_2$ \cite{kozawa2014photocarrier}. The integrated intensity is consistently lower than this maximum at other
excitation energies, including those corresponding to resonances in MoS$_2$ (see for example the weak peak
corresponding to MoS$_2$ X$_{\text{A}}$, appearing as a shoulder of the main PLE peak in Fig.\,\ref{fig:DOP_PLE}(a)). This is
the exact opposite of what is observed in a heterostructure system where energy transfer has been demonstrated. In such system,
the emission intensity of one material is significantly enhanced when the excitation energy is resonant with excitonic
transitions of the other material \cite{kozawa2016evidence}. This consideration allows us to safely rule out energy transfer
between the different layers.

To investigate whether the trilayer stack shows charge transfer, we initially consider the normalized intensity of exciton and
trion when the excitation energy is resonant with excitonic transitions of MoS$_2$. For resonances with both A and B exciton of
MoS$_2$, the emission intensity is enhanced more for MoSe$_2$ A exciton than for the trion. This is consistent with the presence
of a large n-type background doping (free electrons) in CVD-grown MoSe$_2$, which gives rise to strong charged exciton emission
even in the absence of gating (see Fig.\,\ref{fig:TrilayerPL}(a)). For a MoS$_2$/MoSe$_2$ heterojunction, the band alignment
promotes the transfer of holes from MoS$_2$ to MoSe$_2$ \cite{kang2013band}. When we optically excite the trilayer stack, we
induce a net transfer of holes from MoS$_2$ to MoSe$_2$, resulting in a relatively stronger emission of the neutral exciton as
compared to the trion \cite{kozawa2016evidence}.

We also performed circular polarization resolved PLE focusing on MoSe$_2$ incorporated in a trilayer stack. When the excitation
energy is far from resonance (A or B excitons of both materials), MoSe$_2$ shows an extremely small valley polarization. This is
illustrated by the polarization-resolved $\mu$PL spectra of Fig.\,\ref{fig:DOP_PLE}(b), where a negligibly small DCP, defined as
$\text{DCP}=[I_{\sigma_1}-I_{\sigma_2}]/[I_{\sigma_1}+I_{\sigma_2}]$, is observed across the entire energy range of interest
\cite{wang2015polarization}. When the excitation energy of the laser was tuned to the proximity of the resonance with the A
exciton of MoS$_2$, a significant valley polarization accumulates. An example of polarization resolved spectra at an excitation
energy resonant with A exciton of MoS$_2$ is shown in Fig.\,\ref{fig:DOP_PLE}(c). The valley polarization is quantified by a
slightly positive DCP at energies around A exciton and trion of MoSe$_2$ (see Fig.\,\ref{fig:DOP_PLE}(c)). In
Fig.\,\ref{fig:DOP_PLE}(d) we illustrate the excitation energy dependence of the integrated DCP (estimated by extracting the
relevant integrated intensity of exciton and trion with Gaussian fits). We notice a significant increase of the integrated DCP at
energies corresponding to the A exciton resonance of MoS$_2$ (see dashed line in Fig.\ \ref{fig:DOP_PLE}(d)),
with a low energy shoulder possibly related to an enhanced DCP at excitation energies corresponding to MoSe$_2$ B exciton
\cite{wang2015polarization}. Polarization resolved electroluminescence of single and multilayer MoSe$_2$ \cite{onga2016high} as
well as polarization resolved PL of  indirect excitons emitted by a WSe$_2$/MoSe$_2$ heterostructure \cite{rivera2016valley} have
demonstrated higher degree of circular polarization. However, the mechanisms leading to polarized emission in these systems are
fundamentally different from those yielding polarized PL of MoSe$_2$ \cite{onga2016high}, which is virtually impossible to
achieve unless the PL is excited using quasi-resonant excitation in X$_{\text{A}}$ of MoSe$_2$
\cite{wang2015polarization,kioseoglou2016optical,baranowski2017dark}.

We ascribe the observed MoSe$_2$ valley polarization to the hole transfer from MoS$_2$, with a mechanism schematically
illustrated in Fig.\,\ref{fig:DOP_PLE}(e). We excite the trilayer stack with circularly polarized light with a given helicity and
in resonance with the A exciton of MoS$_2$. The valley polarization directly created in MoSe$_2$ is quickly lost, which results
in a negligible DCP, similarly to non-resonant excitation (see Fig.\,\ref{fig:DOP_PLE}(d)). Resonant excitation in MoS$_2$
creates valley polarization for a duration estimated in the hundreds of femtoseconds range \cite{mai2013many}. Charge transfer in
van der Waals heterostructures is an ultra fast process, with upper bounds in the tens of femtoseconds range (hole transfer from
MoS$_2$ has been reported to be faster than \SI{50}{\femto\s} \cite{hong2014ultrafast}). During this very rapid transfer, we
assume that the hole spin (and thus valley due to the large spin orbit splitting in the valence band) is conserved, and, owing to
the excess electron population in the MoSe$_2$ layer, the injected hole forms an exciton populating the valley corresponding to
the helicity of the incoming light. These excitons have presumably a low kinetic energy (no excess energy of the photocreated
hole), which slows down significantly the inter-valley scattering rate due to electron-hole exchange interaction
\cite{yu2014valley}. As a result, this hole transfer is responsible for the observed valley polarization.

\section{Conclusions}

A detailed investigation of the optical properties of MoS$_2$/MoSe$_2$/MoS$_2$ trilayers reveals that stacking dramatically
improves the optical quality of CVD-grown MoSe$_2$, essentially eliminating all defect bound exciton and
MoS$_2$-related emission. These results open the way to using CVD-grown TMDCs for applications and studies that require materials
with excellent optical quality. Photoluminescence spectra from MoSe$_2$ in a trilayer stack are dominated by narrow neutral and
charged A exciton emission, resembling the spectrum of a high quality mechanically exfoliated flake. Density functional
calculations confirm a defect healing scenario in which S atoms replace Se vacancies. Circular polarization resolved PL
measurements demonstrate that MoSe$_2$ exhibits a significant valley polarization even when the excitation energy is far from the
A exciton resonance. This behavior is the signature of an efficient spin-conserving hole transfer from MoS$_2$ to MoSe$_2$. Our
approach provides a robust and straightforward method of healing defects in CVD-grown samples, which might also be beneficial for
the transport properties of these materials.

\section{Acknowledgments}
This work was partially supported by ANR JCJC project milliPICS, the
R\'egion Midi-Pyr\'en\'ees under contract MESR 13053031, BLAPHENE
project under IDEX program Emergence, ``Programme des Investissements d'Avenir'' under the program ANR-11-IDEX-0002-02, reference ANR-10-LABX-0037-NEXT, Swiss
SNF Sinergia Grant no.\ 147607, the European Commission (ITN MoWSeS,
GA 317451), and Deutsche Forschungsgemeinschaft (HE 3543/27-1 and
GRK 2247/1 (QM3)). Y.J., A.K.\ and T.H.\ thank ZIH Dresden for
supercomputer time. This project has received funding from the European Union's Horizon 2020 research and innovation programme under grant agreement No.\ 696656 (Graphene Flagship).

\begin{suppinfo}

Supporting information: Low temperature \si{\micro}PL spectra, comparison of \si{\micro}PL spectra of as-grown versus trilayer
samples, temperature dependent \si{\micro}PL spectra and time-resolved PL.

\end{suppinfo}


\section{Methods}

\subsection{Sample preparation}

The sample with MoS$_2$/MoSe$_2$/MoS$_2$ trilayer stacks, schematically shown in Fig.\,\ref{fig:TrilayerPL}(a), was obtained by
two separate transfer steps using a wet transfer KOH method \cite{wang2016interlayer}. First of all, the upper-MoS$_2$ was
transferred onto the as-grown MoSe$_2$ monlayer on sapphire \cite{dumcenco2015large,mitioglu2016magnetoexcitons}. Subsequently,
the MoS$_2$/MoSe$_2$ stack was transferred onto an as-grown bottom MoS$_2$ monolayer on sapphire. For both transfers, sapphire
chips with material (upper-MoS$_2$ or MoS$_2$/MoSe$_2$ stack) were first spin coated with PMMA 950 at 1500 rpm for \SI{60}{\s}
and baked at \SI{180}{\celsius} for 5 minutes. The films were detached in KOH (30\%) at moderate temperatures
(\SI{70}{\celsius}), washed 3 times in deionized water, transferred onto sapphire with the stacking layer (MoSe$_2$ or
bottom-MoS$_2$) and dried at \SI{50}{\celsius}  for 30 minutes. The PMMA was removed by dipping the sample in acetone for 12
hours, followed by rinsing with isopropanol and drying in a N$_2$ flow. Such method provides a polymer clean interface of stacks
and minimal damage of material caused by the transfer process. Using this method, a large area film with monolayer (MoS$_2$),
bilayer (MoSe$_2$/MoS$_2$, MoS$_2$/MoS$_2$ and MoS$_2$/MoSe$_2$ stacks) and trilayer (MoS$_2$/MoSe$_2$/MoS$_2$ stack) was
obtained. A micrograph of a representative transfer area is shown in Fig.\,\ref{fig:TrilayerPL}(b). To determine the position of
areas with different number of stacked layers, Ni markers were deposited.

\subsection{Optical measurements}

For the optical characterization, the sapphire substrate was mounted on the cold finger of a He-flow croystat. The excitation was
provided either by a CW frequency doubled solid state laser emitting at \SI{532}{\nano\m} or by the frequency doubled output of
an optical parametric oscillator (OPO), synchronously pumped by a mode-locked Ti:sapphire laser. The typical temporal pulse width
was \SI{300}{\femto\s}, with a repetition rate of \SI{80}{\mega\Hz}. The excitation beam was focused on the sample by a
$50\times$ microscope objective, giving a spot size of approximately \SI{1}{\micro\m} and having a numerical aperture of 0.55.
The emitted PL was collected through the same objective and redirected to a spectrometer equipped with a liquid nitrogen cooled
CCD camera or (for time-resolved measurements) to an imagining spectrometer and detected using a synchroscan streak camera with
the temporal resolution set to \SI{5}{\pico\s}. All the spectra have been measured at \SI{5}{\K}, unless
otherwise specified.

For spatial mapping the emission has been monitored while the
optical cryostat was displaced with respect to the microscope
objective using high precision motorized $x-y$ translation stages
(\SI{1}{\micro\m} step). The integrated intensity of given features
(\emph{e.g.} A exciton emission) has been obtained by performing
Gaussian fitting of the measured \si{\micro}PL spectra.

\subsection{Density functional theory of band structure in MoSe$_2$/MoS$_2$ heterobilayers}
We have calculated heterobilayers made of MoSe$_2$ and MoS$_2$ monolayers, using DFT as implemented in the Crystal09
software\cite{Crystal09}. We employed all-electron Gaussian-type bases of
triple-quality\cite{Peintinger2013,lichanot1993quantum}, while Mo atoms were treated with the HAYWSC-311(d31)G basis with
effective core potential set\cite{Cora1997}, together with the PBE gradient corrected density functional\cite{Perdew1996}.
London-dispersion interactions were accounted for using the approach proposed by Grimme (DFT-D3)\cite{Grimme2006}. Full
optimization of atomic positions and lattice vectors was performed on both models: Model 1 with perfect MoS$_2$ monolayer and one
Se vacancy in the MoSe$_2$ monolayer, and Model 2 with one S vacancy in the MoS$_2$ monolayer and one S substitution in the
MoSe$_2$ monolayer. The models are built of $5 \times 5$ supercells (see Fig.\,\ref{fig:DFTbandstructure}).

Geometry optimization only slightly alters the lattice geometry (see Table \ref{tab:lattice}).
\begin{table}[h!]
\caption{Structural parameters of MoS$_2$/MoSe$_2$ defective
heterostructures giving the lattice constant ($a$) and interlayer
metal-to-metal distances ($d$).}\label{tab:lattice}
\begin{center}
\begin{tabular}{l|c|c|}

System & $a$ (\si{\angstrom}) & $d$ (\si{\angstrom})\\
\hline MoS$_2$ 1L & 3.171 &  - \\
MoSe$_2$ 1L & 3.251 &  - \\
MoS$_2$/MoSe$_2$ & 3.214 & 6.21 \\
Model 1 & 3.206 & 6.30 \\
Model 2 & 3.203 &  6.26\\
\end{tabular}
\end{center}
\end{table}

For the large supercells, which still overestimate the defect density in experiment, the defects studied here do not introduce
any drastic changes into the structural properties of the systems. However, we note that we are constrained with the commensurate
models of a heterobilayer, in which the corresponding monolayers are slightly distorted compared with the relaxed monolayers
\SI{3.251}{\angstrom} and \SI{3.171}{\angstrom} for MoSe$_2$ and MoS$_2$, respectively. For a perfect heterobilayer
(\SI{3.214}{\angstrom}), this gives $1.15\%$ compression of MoSe$_2$ and $1.4\%$ elongation of MoS$_2$.

\bibliography{BibliographyTrilayers}

\end{document}